\documentclass{natureprintstyle}

\usepackage{graphicx}

\newcommand{\apj}{Astrophys. J.}
\newcommand{\apjl}{Astrophys. J.}
\newcommand{\mnras}{Mon. Not. R. Astron. Soc.}
\newcommand{\nat}{Nature}
\newcommand{\aap}{Astron. Astrophys.}

\title{Monster black holes}

\author{\bf MICHELE CAPPELLARI}

\begin{document}

\maketitle

\begin{abstract}
A combination of ground-based and spacecraft observations has uncovered two black holes of 10 billion solar masses in the nearby Universe. The finding sheds light on how these cosmic monsters co-evolve with galaxies.
\end{abstract}

Giant black holes, with masses of a few billion times that of the Sun, have fascinated scientists, science-fiction writers and the general public alike since they were first proposed\cite{Lynden-Bell1969} four decades ago. These supermassive black holes, which lurk in the centre of galaxies, are not just theoretical curiosities. Their existence was convincingly demonstrated 15 years ago, and soon astrophysicists realized that such black holes can have profound effects on how galaxies form\cite{Silk1998}. This realization spurred a flurry of studies aiming to understand the joint evolution of galaxies and black holes. On page 215 of this issue, McConnell et al.\cite{McConnell2011} present the detection of the two most massive black holes ever found. They argue that their finding provides a key missing piece of evidence to our understanding of how galaxies and black holes form.

The existence of monster black holes was originally invoked to explain the intense energy released by active galaxies known as quasars. These galaxies are especially numerous at very large distances from Earth, corresponding to a time when the Universe was less than half its current age. Their tremendous brightness is thought to be the last flash from gas accelerated to extreme speed before being accreted by a giant black hole in the galaxy's nucleus. There is now less gas in the nearby Universe than there was at early epochs, because most of it was used to make stars. For this reason, no quasar exists in our cosmic backyard. However, if supermassive black holes are responsible for quasars' power production, they should still lie dormant in the centre of the most massive nearby galaxies.

Indeed, systematic searches for supermassive black holes have found them at the centre of all massive galaxies for which reliable determinations of the black-hole mass exist (see McConnell et al.\cite{McConnell2011} for an up-to-date list). But such black holes are not massive enough to power the brightest quasars. The ideal candidate galaxies for hosting these largest black holes are the massive elliptical galaxies found at the centre of galaxy clusters. Like giant spiders at the centre of a web, these galaxies lie at the bottom of a cluster's gravitational potential well, feeding their black hole by accreting gas and stars from neighbouring galaxies. However, the nearest big galaxy cluster to Earth is about 100 megaparsecs away, and this large distance makes determining the mass of their putative black holes a challenge.

McConnell et al.\cite{McConnell2011} targeted the central elliptical galaxy of two massive clusters, using two of the world's largest telescopes and observations from the Hubble Space Telescope. They found two black holes, each with a mass of more than 10 billion solar masses, in the nuclei of the two galaxies. These objects probably represent the missing dormant relics of the giant black holes that powered the brightest quasars in the early Universe.

The mass of the supermassive black holes studied so far is closely related to the amount of random motion --- the velocity dispersion --- of the stars in the central parts of the host galaxies\cite{Ferrarese2000,Gebhardt2000bh}. The velocity dispersion is related to the mass of the spheroidal component, the stellar bulge, of the galaxies. The existence of this empirical relationship is attributed to the fact that the black hole grows by feeding from the gas that is accreted by its host galaxy. This gas accretion leads to the simultaneous growth of both the black hole and the galaxy stellar bulge, until the intense energy produced by the black hole heats up or blows away the gas, halting both star formation and black-hole accretion\cite{DiMatteo2005} (Fig.~1a).

\begin{figure*}
\begin{center}
\includegraphics[width=0.8\textwidth]{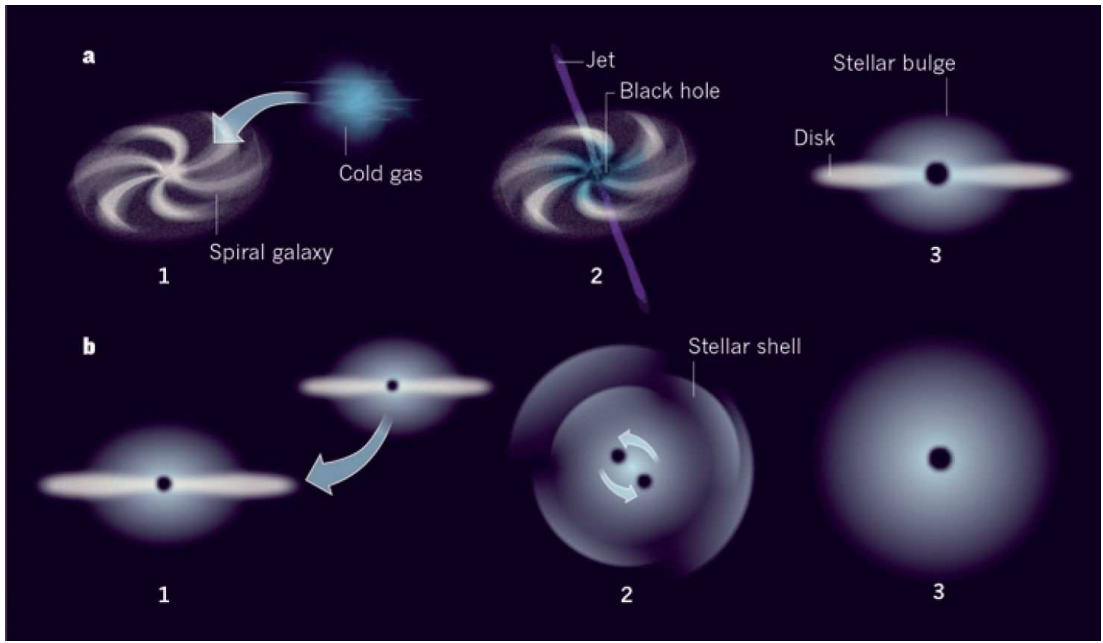}
\end{center}
\caption{{\bf Two ways to grow galaxies and their central black holes. a,}  In the gas-accretion hypothesis, a pure-disk, gas-rich spiral galaxy, possibly with a small black hole, accretes cold gas either from the environment (1) or by merging with another spiral galaxy (not shown). The gas sinks to the galaxy's centre, forming a black hole and a stellar bulge. This process terminates when the galaxy becomes 'active' and produces two opposing high-speed jets of particles (2). The end product is a lenticular galaxy, which has a supermassive black hole, disk and stellar bulge, and larger stellar velocity dispersion than the initial galaxy (3). {\bf b,} Alternatively, the galaxy and black hole can grow by 'dry' merging: a gas-poor lenticular galaxy accretes a similar galaxy (1). The two galaxies merge and form stellar shells, while their black holes spiral towards the centre until they coalesce (2). The result is a spheroidal galaxy that has a larger size and bigger black hole than the progenitors, but nearly unchanged velocity dispersion (3). McConnell and colleagues' analysis\cite{McConnell2011} indicates that the second route may dominate the growth of the most massive black holes.}
\end{figure*}

Interestingly, the two newly measured supermassive black holes are more massive than would be predicted from their velocity dispersion. This suggests that, unlike their smaller counterparts, these black holes did not grow most of their mass by gas accretion but instead grew by the 'dry' merging of gas-poor galaxies. In a dry merger of two similar galaxies, the resulting galaxy's stellar and black-hole masses are the sums of the corresponding masses of the two progenitor galaxies (Fig.~1b). However, contrary to the gas-accretion case, the velocity dispersion of the final galaxy's stars remains unchanged, or changes little if the mass doubling is achieved through a sequence of several smaller mergers of gas-poor galaxies\cite{Naab2009}. Either way, the merging process leaves behind a black hole that is more massive than would be predicted from the mass-velocity dispersion relationship --- just like the two black holes observed by McConnell and colleagues\cite{McConnell2011}.

Not everybody agrees that the observed mass–velocity dispersion relationship indicates that black holes regulate galaxy growth by stopping gas accretion with their energy production (Fig.~1a). Studies\cite{Peng2007,Jahnke2011} have shown that starting from a set of galaxies containing black holes of random masses, a sequence of mergers can lead to a relationship close to the observed one without the need for energy input from the black hole. However, this simulated relationship deviates systematically from the observed one. These claims\cite{Peng2007,Jahnke2011} can be tested by accurately measuring black-hole masses in galaxies with different properties, as McConnell et al. did. Progress requires the largest telescopes and the use of state-of-the-art technologies such as integral-field spectroscopy, to map the motion of the stars in two dimensions, and adaptive optics, to correct the blurring effect of Earth's atmosphere and achieve sharp images\cite{Cappellari2009cena}. The future looks bright for black-hole studies using the next generation of 40-metre telescopes, such as the European Extremely Large Telescope, which will significantly increase the number of galaxies that can be reliably investigated.

\begin{addendum}
 \item[Michele Cappellari] is in the Sub-department of Astrophysics,\\ Department of Physics, University of Oxford, Oxford OX1 3RH, UK.\\ e-mail: cappellari@astro.ox.ac.uk
\end{addendum}

\end{document}